\documentstyle[pre,aps,amsfonts]{revtex}

\input epsf

\begin{document}
\draft
\twocolumn[\hsize\textwidth\columnwidth\hsize\csname@twocolumnfalse\endcsname
\title{Stability of Texture and Shape of Circular Domains of Langmuir
Monolayers} 
\author{David Pettey and T.C. Lubensky}
\address{Department of Physics and Astronomy, University of Pennsylvania,
Philadelphia, PA 19104, USA}
\maketitle

\begin{abstract}
Finite domains of a Langmuir monolayer in a phase with tilted
molecules can be modeled by a simple elastic free energy of an $XY$
order parameter with isotropic and anisotropic line tension terms.
The domains can and often do contain nontrivial textures, which in
turn influence the shape of the domains.  Herein we investigate the
properties of a simplified isotropic model with a single elastic
constant.  For circular domains a first-order phase transition is
found between two distinct textures: an exterior defect (or ``virtual
boojum'') texture, and an interior defect texture.  Starting with a
circular domain and either of these two textures as a ground state, we
find shape instabilities develop that depend on the elastic constants
and line tensions in the simplified model.  In both cases a necessary
but not sufficient condition for the onset of shape instabilities is
the possibility for a local negative effective line tension to develop
from the anisotropic line tension term.
\end{abstract}

% Langmuir-Blodgett films
% Thin film structure and morphology
% Theory and models of LC structure.
% Defects in LC's
% Defect and impurities: doping, implantation, distribution,
%  concentration, etc.
% Physical properties of thin films, nonelectric.

\vskip2pc]

\section{Introduction} \label{sn:Introduction}
\par
The biological importance of thin films of surfactants, such as DPPC
(a primary component of human lung surfactant \cite{biology}), and
other potential commercial applications of self-assembled structures
\cite{Roberts} has spurred a resurgence of research in the study of
Langmuir films.  For the physicist, monolayers of surfactants at the
air-water interface provide an interesting $2D$ system with very rich
phase behavior \cite{Knobler92,Mohwald90}. Techniques such as
fluorescence microscopy and Brewster angle microscopy allow direct
observation of the system as it proceeds through a phase transition.
X-ray scattering can also be used.  Atomic force microscopy on
Langmuir-Blodgett films can provide more detailed information that
appears to be consistent with passive observations of the Langmuir
film before extraction \cite{Yang94}.
\par
Typically, as a film is compressed, it passes from a gas phase (G) to
a more condensed liquid expanded phase (LE) to an even more ordered
liquid condensed phase (LC) and then, sometimes, into a solid phase
before finally becoming a multi-layer film.  Experiments have revealed
a very complex variety of patterns, shapes, and textures present in
the LC/LE and LE/G coexistence regions \cite{Weis84,Knobler90}.  Upon
rapid compression of the film, highly branched structures are often
seen.  These appear to be the result of diffusion limited aggregation
\cite{Knobler90}, and typically they relax to more regular shapes.
More modest compression rates tend to produce a finite number of
condensed domains that grow in size, but not number \cite{Helm88}, as
the film is compressed through a coexistence region.  Films composed
of enantiomeric surfactants sometimes exhibit domains with a preferred
handedness (chirality) to their shape.  The handedness alternates with
the choice of enantiomer and in racemic mixtures the domains are
achiral \cite{Weis84,Weis85,Yang94}.  Within domains of the ordered
phase, complex structures can sometimes be observed as well
\cite{Qiu91,Riviere95}.  The addition of cholesterol to these systems
often leads to labyrinthine or striped patterns, and to spiral domains
of a preferred handedness \cite{Weis85}.
\par
Molecules in LC phases of Langmuir films either align normal to the
air-water interface, or they can tilt relative to the interface
\cite{Zhai97,Rasing85}, thereby defining a two-dimensional vector in
the plane of the film.  In addition some films exhibit hexatic order
as well as tilt order \cite{Hui93,Smith90}.  Domains of both tilted
and non-tilted phases exhibit non-circular shapes
\cite{Weis84,Riviere95,Weis85,Yang94,Bruinsma}.  Static and dynamic
properties of non-tilted domains as well as modulated equilibrium
phases are well explained by dipole interactions among the electric
dipoles aligned normal to the layers
\cite{Andelman87,Vanderlick90,McConnell90,Koker93,Goldstein,Seul,Lee95,McConnell88}.
Domain shapes and textures in tilted domains clearly depend on the
existence of orientational order in the plane of the film
\cite{Riviere95,Knobler}.  For example, it is difficult to explain
chiral shapes that appear in chiral films by a dipole model.
\par
In this paper, we will consider a particularly simple model for tilted
domains in which tilt order is described by an isotropic $XY$-model
and in which coupling between texture and domain shape arises because
of an interaction favoring alignment of $XY$-vectors at a specific
angle with respect to the domain boundaries.  This model is clearly
oversimplified, but it can explain observed shapes and textures in
some systems \cite{Riviere95,Bruinsma,Knobler}.  A complete model
would allow for different elastic constants for splay and bend
distortions and for splay-bend coupling in chiral systems \cite{Kraus}
in addition to interactions between electric dipoles with components
both parallel and perpendicular to the film.  Nonetheless, this
simplified isotropic model predicts non-trivial shapes and textures,
the full range of which have yet to be analyzed.  In spite of its
simplicity, it involves highly non-trivial couplings between texture
and shape that need to be understood before more realistic models can
be studied.
\par
Our model is characterized by three energy parameters; the isotropic
elastic constant $K$ (units of energy), the isotropic line tension
$\gamma$ (units of energy/length) and the line tension $\mu$ favoring
alignment of the $XY$-vector ${\bf \hat{m}}$ with the tangent to the
domain perimeter ${\bf \hat{t}}$.  We obtain the following results for
this model.  For $\mu/\gamma < 1$, circular domains are globally
stable.  Circular domains of radius $R$ can have two equilibrium
textures: one with a strength $+1$ disclination \cite{Lubensky} with
core radius $\xi$ at the center of the domain and one described by a
$+2$ defect (virtual boojum) exterior to the domain \cite{Sethna}.
The interior defect texture is favored for large values of $R/\xi$
whereas the exterior defect texture is favored for large values of
$K/\mu\xi$.  For a circular domain, the exterior defect texture never
becomes locally unstable.  The interior defect texture does, however,
become locally unstable with respect to moving the defect toward the
edge of the circle.  When $\mu/\gamma>1$, the circular shape and
associated textures can become unstable.  When the defect is in the
interior, the circle can become unstable with respect to $n$-fold
modulations as the domain radius $R$ is increased.  The value of $n$
that first becomes unstable depends on $\mu/\gamma$.  When the defect
is outside, the domain becomes unstable with respect to $n=2$
distortions first, as $R$ is increased.
\par
We will proceed with our investigations as follows.  We will review
the interior and exterior defect textures for a circular domain and
then, comparing them, will present a phase diagram for these two
textures.  Next, we will examine the shape stability by taking an
initial configuration consisting of a circular domain and one of the
two extremal textures.  For each of the two textures we will then
allow for perturbations in the shape and texture, calculate an
effective free energy dependent only upon the shape, and examine the
stability of the circular domain with respect to shape deformations.

\section{The model and our aims} \label{sn:model}
\par
Henceforth we will be concerned only with the following problem.  Find
the minimum of the free energy \cite{fn:chiral_term}
\begin{eqnarray}
F &=& \frac{1}{2} \int_D K_s \left[ \left(\nabla \cdot {\bf \hat{m}}' \right)^2  +
K_b\left(\nabla \times {\bf \hat{m}}' \right)^2 \right] d^2x \nonumber\\
&&-\mu' \oint_{\partial D} {\bf \hat{m}}' \cdot {\bf \hat{t}} ds 
- \eta' \oint_{\partial D} {\bf \hat{m}}' \times {\bf \hat{t}} ds   
+ \gamma \oint_{\partial D} ds,
\end{eqnarray}
over all domains $D$ with fixed area $A$ ($=\pi R^2$), and over all unit vector
fields ${\bf \hat{m}}'$, for given values of the elastic constants $(K_s, K_b,
\mu', \eta', \gamma)$.  $K_s$ and $K_b$ are the usual bend and splay
elastic constants, $\gamma$ is a line tension, ${\bf \hat{t}}$ is the tangent
vector to the curve $\partial D$, and $\mu'$ and $\eta'$ are the
coefficients of spontaneous bend and splay, respectively.  We have
chosen to write the spontaneous bend and splay contributions as line
integrals using
\begin{eqnarray}
\int_D \nabla \cdot {\bf \hat{m}}' d^2x &=& \oint_{\partial D} {\bf \hat{m}}' \times
{\bf \hat{t}} ds \nonumber \\
\int_D  \nabla \times {\bf \hat{m}}' d^2x &=& \oint_{\partial D} {\bf \hat{m}}' \cdot
{\bf \hat{t}} ds.
\end{eqnarray}
This allows us to identify $-\mu' {\bf \hat{m}}' \cdot {\bf \hat{t}} -
\eta' {\bf \hat{m}}' \times {\bf \hat{t}} + \gamma$ as an effective
anisotropic line tension.  If $\gamma < \sqrt{\mu^{'2} + \eta^{'2}}$,
then choosing ${\bf \hat{m}}'$ to follow ${\bf \hat{t}}$ for a given
curve $\partial D$ will allow the anisotropic line tension to be
negative.  We will see that within the context of the
one-coupling-constant approximation $(K=K_s=K_b)$, allowing the
effective line tension to become negative will be a necessary
condition for instabilities in the shape of a circular domain.
\par
We will now restrict our attention to the one-coupling-constant
approximation, simplifying the form of the bulk free energy.  Defining
$\Phi'$ via
\begin{equation}
{\bf \hat{m}}' = \cos\Phi' {\bf \hat{e}}_x + \sin\Phi' {\bf \hat{e}}_y,
\end{equation}
we have
\begin{eqnarray}
&&\int_D \left[ K_s \left(\nabla \cdot {\bf \hat{m}}'\right)^2  +
K_b\left(\nabla \times {\bf \hat{m}}'\right)^2 \right] d^2x \nonumber \\
&=& K \int_D \nabla \Phi' \cdot \nabla \Phi' d^2x.
\end{eqnarray}
In appendix \ref{nochiralapp} we show that the transformation
\begin{eqnarray}
&\Phi& = \Phi' - \tan^{-1}\frac{\mu'}{\eta'} - \frac{\pi}{2} \nonumber\\
&\mu& = \sqrt{\mu^{'2} + \eta^{'2}} \label{transformation} \\
&{\bf \hat{m}}& = \cos\Phi {\bf \hat{e}}_x + \sin\Phi{\bf \hat{e}}_y \nonumber
\end{eqnarray}
simplifies the free energy further:
\begin{equation}
F = \frac{1}{2} K \int_D (\nabla \Phi)^2 d^2x
-\mu\oint_{\partial D} {\bf \hat{m}} \cdot {\bf \hat{t}} ds 
+ \gamma \oint_{\partial D} ds.
\label{Fmu}
\end{equation}
We could just as easily have transformed away the spontaneous bend
term in favor of an effective coefficient for spontaneous splay.
Thus, although Eq.(\ref{Fmu}) is a chiral free energy (the spontaneous
bend term is chiral), chirality can only manifest itself through
the texture ${\bf \hat{m}}$ and not through the domain shape $D$ ($D$ is
invariant under the transformation, see App. \ref{nochiralapp}).
\par
To make contact with previous work \cite{Bruinsma} we note that we
could have written our surface term in the form
\begin{equation}
\oint_{\partial D} \sigma(\theta-\Phi) ds,
\end{equation}
where the outward normal to the boundary is given by
\begin{equation}
{\bf \hat{n}} = \cos\theta {\bf \hat{e}}_x + \sin\theta {\bf \hat{e}}_y,
\end{equation}
and 
\begin{equation}
\sigma(x) = \gamma + \sum_{n=1}^{\infty} \left( a_n \cos nx +b_n \sin
nx \right).
\end{equation}
Taking $b_1=-\mu$ with all other $a_n$'s and $b_n$'s set to zero will
recover our effective line tension
\begin{equation}
\gamma\oint_{\partial D} ds - \mu \oint_{\partial D} {\bf \hat{m}} \cdot {\bf \hat{t}}
ds.
\end{equation}
Similarly a nonzero $a_1$ will produce a spontaneous splay
contribution.  The transformation (\ref{transformation}), in this
language, allows the elimination of the $a_1$ ($b_1$) term in favor of
a new effective $b_1'$ ($a_1'$).  
\par
Rudnick and Bruinsma \cite{Bruinsma} have demonstrated that a
nonzero $a_2$, in addition to a nonzero $a_1$, leads to non-circular
shapes and nontrivial textures in agreement with some experiments
\cite{Knobler}.  However, Rivi\`{e}re and Meunier \cite{Riviere95} can
account for these same shapes and textures without a nonzero $a_2$ by
incorporating the apparent anisotropy between $K_s$ and $K_b$.  We
will show that even with $K_s=K_b$ and $a_2=0$ [as in Eq. (\ref{Fmu})]
shape instabilities {\em still} exist when $\gamma<\mu$, that is, once
a negative effective line tension is allowed.
\par
To see that instabilities should arise, we first identify
\begin{eqnarray}
&&A_b = \frac{K}{\mu R} \nonumber\\
&&A_s = \frac{\mu}{\gamma} -1
\end{eqnarray}
as two independent dimensionless parameters.  $A_b$ is a measure of
the competition between the bulk energy and the anisotropic piece of
the line tension.  For a given $\partial D$, $A_b$ will determine the
preferred texture.  $A_s$ measures the relative strength of the
anisotropic and isotropic contributions to the effective line tension.
When $A_s>0$, the effective line tension can be negative.
\par
When $A_b \rightarrow \infty$ we can assume that $K \rightarrow
\infty$ and hence that ${\bf \hat{m}}$ will be constrained to be a constant.
Thus, without loss of generality, we can assume that ${\bf \hat{m}}={\bf \hat{e}}_y$.
This form for ${\bf \hat{m}}$ yields
\begin{equation}
\oint_{\partial D} {\bf \hat{m}} \cdot {\bf \hat{t}} ds = 0
\end{equation}
for all $\partial D$, and as such the value of $\mu$ is irrelevant.
In fact, the free energy is simply
\begin{equation}
F = \gamma \oint_{\partial D} ds.
\end{equation}
For fixed area, $A$, the preferred shape will simply be a circle.
\par
In contrast, when $A_b \rightarrow 0$ we assume $K \rightarrow 0$ and
thus ${\bf \hat{m}}$ can take any form it chooses in the bulk.  In particular,
we can always choose ${\bf \hat{m}}$ such that ${\bf \hat{m}} \cdot {\bf \hat{t}}=1$ at all
points on the curve, for any curve $\partial D$.  Thus now our free
energy becomes
\begin{equation}
F = (\gamma - \mu) \oint_{\partial D} ds.
\end{equation}
If $\gamma>\mu$ then certainly we again have a circle as the preferred
shape.  But if $\gamma< \mu$ then the domain attempts to maximize the
length of its boundary, $\partial D$.  This results in pathologically
distorted shapes, or unbounded domains.  Thus, when $\gamma < \mu$ we
anticipate that for large $A_b$ the domain shapes will be circular,
but for small $A_b$ the shapes will begin to distort.  However, when
$\gamma>\mu$ it would not be at all surprising to find that the
circular domains are preferred for all values of $A_b$.
\par
We are now ready to begin our search for optimal shape and texture
configurations.  Although we will restrict our attention to the
one-coupling-constant approximation, which provides us with a
particularly simple bulk Euler-Lagrange equation
\begin{equation}
\nabla^2 \Phi =0,
\end{equation}
it is still not a trivial task to find the optimum $\Phi$ for a
generic domain $D$.  We will approach the problem as follows.  Noting
that experimentally the domains appear to be circular for small $A$,
we will first fix the domain shape to be a circle and try to find the
optimum $\Phi$.  This is tantamount to assuming that $\gamma$ is
infinite.  Having found the optimal texture for a circular domain, we
will then allow the shape and texture to deform.  The next section
will deal with the task of finding the optimal texture for a circular
domain and the succeeding section will deal with instabilities in the
shape of the domain.

\section{The optimal texture for a circular domain} \label{sn:texture}
\par
The main question to address here is whether the domain should contain
a topological defect.  In experiments by Rivi\`{e}re and Meunier
\cite{Riviere95} the domain clearly does not contain a defect and has
the ``virtual boojum'' texture previously described by Sethna
\cite{Sethna}. However in \cite{Qiu91} Qiu {\it et al.} observed a
star defect inside the domain, which indicates that it may be possible
to have an isolated $+1$ defect inside a domain when hexatic order is
not present (as the hexatic order is lost the arms of the star defect
retract and one is left with a simple $+1$ defect
\cite{Meyer,Selinger,LubPettey}).  Also, in thin films of smectic-C
liquid crystals, which can be approximately modeled by the same free
energy, isolated $+1$ defects are often observed \cite{Lubensky}.
\par
Fixing the domain to be a disk, the isotropic line tension term will
be irrelevant to this discussion since it is unaffected by changes in
texture alone.  The anisotropic piece of the boundary energy, however,
favors having ${\bf \hat{m}}$ parallel to the tangent to the boundary and thus
prefers having either a $+1$ defect inside the domain or a $+2$ defect
on the boundary (sometimes referred to as a boojum).  The bulk portion
of the free energy clearly prefers that ${\bf \hat{m}}$ simply be uniform
within the domain.  Thus, whether it is preferable to have a defect
inside the domain or not depends upon the competition between the bulk
elastic energy and the anisotropic line tension, as well as upon the
energy cost required to nucleate a melted defect core.
\par
In this section we will first review the virtual boojum texture, which
is an extremal texture for the case where there is no defect inside
the disk \cite{Bruinsma}.  In fact, in section~\ref{sn:extshape} we
will show that this is a locally minimal texture with respect to our
free energy.  Next we will consider textures associated with the
defect inside the domain.  We will note that the texture for an
isolated $+1$ defect at the center of the domain is extremal and for
$A_b<1$ it is a local minimum as well.  We will investigate a class of
ansatz textures consisting of an isolated $+1$ defect located anywhere
within the domain.  We will find that within this restricted class of
textures, there is a first-order transition between the texture with
the defect at the center of the disk and a texture with the defect
close to the boundary.  Thus, we find that not only is the texture of
the defect at the center of the disk unstable for $A_b>1$ but that it
also fails to be a global minimum for even smaller values of $A_b$
within the class of textures with an enclosed defect.  In contrast,
the virtual boojum texture is a local minimum for all values of $A_b$.
\par
Finally, we compare the virtual boojum texture with the texture of the
defect at the center of the disk.  We find a first-order transition
between these two textures which is most easily described by the
dimensionless parameters $c=K/2\mu\xi$, $x=R/\xi$, and $g$, where
$\xi$ is the coherence length and $g$ is a phenomenological parameter
associated with the energy cost of a defect core (in our calculations
we will take $g=1$).

\subsection{ Virtual Boojum} \label{sn:boojumtexture}

\begin{figure}
\epsfxsize=3.375in
\centerline{\epsfbox{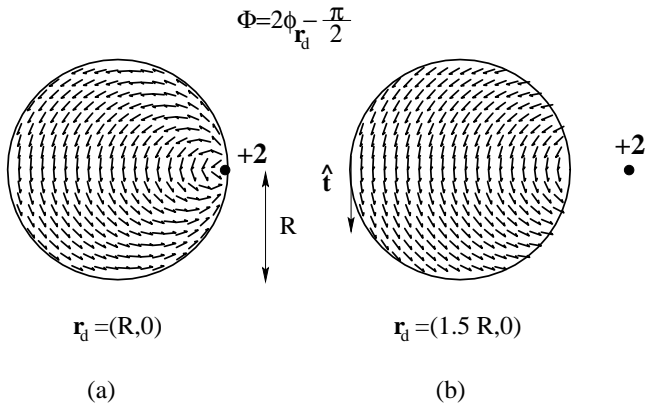}}
\caption{Textures arising from a single $+2$ defect.  The constant term in
$\Phi$ has been chosen to make ${\bf \hat{m}}$ as close to ${\bf \hat{t}}$ (the
tangent vector) as possible.  With the defect at the boundary (a) we
have ${\bf \hat{m}} \cdot {\bf \hat{t}}=1$ but we also have very large gradients in
${\bf \hat{m}}$ near the defect.  In (b) these large gradients have been
expelled along with the defect but now we no longer have ${\bf \hat{m}} \cdot
{\bf \hat{t}}=1$.  The competition between the ideal boundary (${\bf \hat{m}} \cdot
{\bf \hat{t}} =1$) and bulk ($\partial_i {\bf \hat{m}}=0$) conditions leads to a
preferred separation distance between the domain and the defect.
        }
\label{fig_ext_textures}
\end{figure}

\par
Let us now recall the virtual boojum texture \cite{Sethna}.  This is
simply the texture produced by a single $+2$ defect sitting outside
the disk (Fig.\ \ref{fig_ext_textures}).  With the following notation,
\begin{eqnarray}
&&{\bf r} = (r \cos\phi, r \sin\phi) = (x,y) \nonumber\\
&&z = x+iy \\
&&\phi_{{\bf r}_d}({\bf r}) = {\rm Im}(\log|z-z_d|) = \tan^{-1}\left(
\frac{y-y_d}{x-x_d}\right) \nonumber
\end{eqnarray}
the field for a $+2$ defect located at ${\bf r}_d = (r_d\cos\phi_d,
r_d\sin\phi_d)=(x_d,y_d)$ is just
\begin{equation}
\Phi = 2 \phi_{{\bf r}_d}({\bf r}) + C.
\label{exterior_texture}
\end{equation}
Clearly the texture satisfies the bulk Euler-Lagrange equation,
$\nabla^2 \Phi=0$.  Furthermore, this collection of textures (one for
each ${\bf r}_d$ and $C$) smoothly varies in its extremes from the uniform
texture (${\bf r}_d\rightarrow\infty, {\bf \hat{m}}\rightarrow$ constant), which is
preferred when $\mu=0$, to a texture where ${\bf \hat{m}} \cdot {\bf \hat{t}}=1$
($r_d=R$), which is preferred when $\mu\rightarrow\infty$.
\par
Taking the domain to be a disk of radius $R$ centered at the origin we
then find that Eq. (\ref{Fmu}) yields
\begin{eqnarray}
F &=& -2\pi K \log\left( 1-\left(\frac{R}{r_d}\right)^2\right) + 2 \pi
\frac{\mu}{r_d}R^2 \sin (C+\phi_d) \nonumber \\
&&+ 2 \pi \gamma R, \label{F_exterior_phid}
\end{eqnarray}
where we have assumed that the defect is outside the disk at least a
core radius, $\xi$, away from the boundary ($r_d>R+\xi$).
Minimizing over $C$ we find the preferred value of $C$ to be,
\begin{equation}
C^0 = -\frac{\pi}{2} - \phi_d.
\end{equation}
Without loss of generality we can take $\phi_d=0$, and finally write
the energy for a $+2$ defect outside the disk, on the positive
$x-$axis as,
\begin{equation}
F_{+2} = -2\pi K \log\left( 1-\left(\frac{R}{r_d}\right)^2\right) - 2 \pi
\frac{\mu}{r_d}R^2 + 2 \pi \gamma R.
\label{F_exterior}
\end{equation}
For a given value of the dimensionless quantity
\begin{equation}
A_b = \frac{K}{\mu R}
\end{equation}
there is a preferred stable minimum value \cite{Sethna,Bruinsma} for
$r_d$ given by
\begin{equation}
r_d^0 = R(A_b + \sqrt{1+A_b^2}).
\label{ext_position}
\end{equation}
Furthermore, we see that using this value of $r_d$ in our $F_{+2}$
above yields $F_{+2}(r_d^0) < 2\pi\gamma R$, revealing that this
texture is preferred over the uniform texture for all values of $A_b$.
\par
Rudnick and Bruinsma \cite{Bruinsma} have shown that this texture with
the defect at $r_d^0$ is in fact extremal.  We will further show that
this texture is indeed a local minimum when we investigate shape
instabilities in section~\ref{sn:shapechanges}.

\begin{figure}[t]
\twocolumn[\hsize\textwidth\columnwidth\hsize\csname@twocolumnfalse\endcsname
\epsfxsize=7in
\centerline{\epsfbox{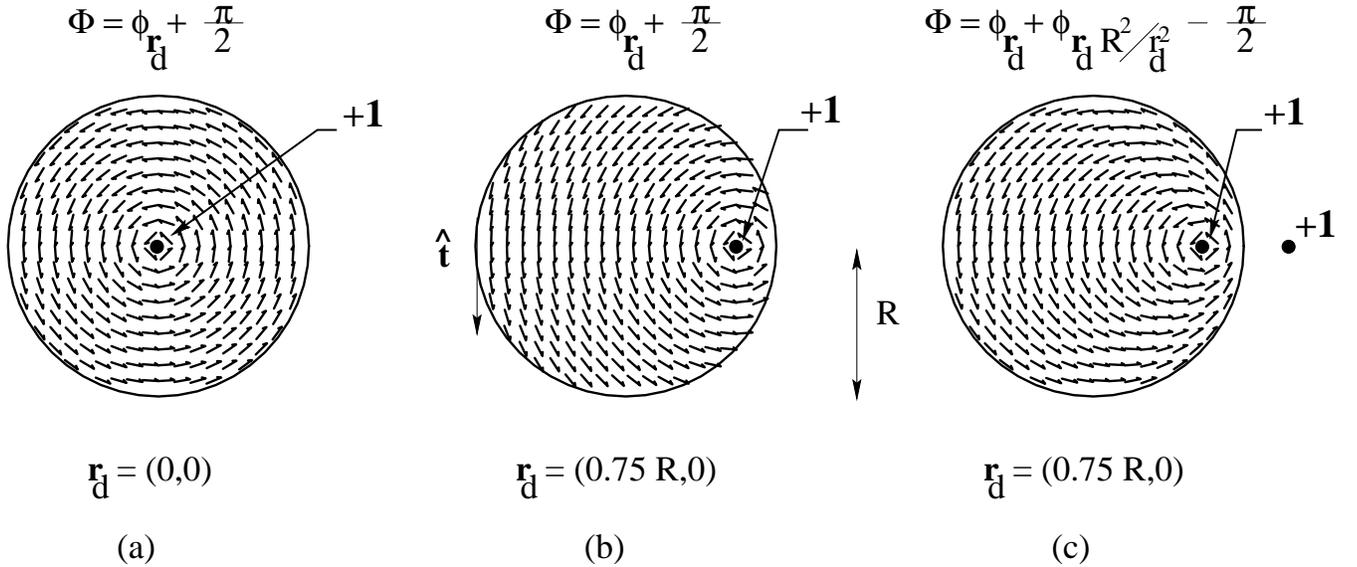}}
\caption{Texture arising from a $+1$ defect inside the disk.  The textures in
(a) and (b) arise solely from the interior defect, whereas the texture
in (c) includes the effects of an image defect outside the domain.
Notice that in (a) and (b) we have ${\bf \hat{m}} \cdot {\bf \hat{t}}=1$ whereas in
(b) we do not. }
\label{fig_int_textures}
\vskip 2pc]
\end{figure}

\subsection{Defect Inside} \label{sn:defecttexture}
\par
Here we will look at the class of ansatz textures
\begin{equation}
\Phi = \phi_{{\bf r}_d}({\bf r}) + \frac{\pi}{2}
\label{interior_texture}
\end{equation}
where we assume $r_d<R-\xi$ (i.e. the defect is inside the disk and at
least a core radius, $\xi$, away from the boundary).  Without loss of
generality we take $\phi_d=0$ and accordingly we have already chosen
the constant term to be the optimal value.  Again, clearly, all such
textures satisfy the bulk Euler-Lagrange equation, $\nabla^2\Phi=0$,
within the domain, except at the position of the defect.  We note that
when $r_d=0$, ${\bf \hat{m}}$ is parallel to the tangent to the boundary
(Fig.\ \ref{fig_int_textures}).
\par
Applying Eq. (\ref{Fmu}) we find that the energy of this texture is
\begin{eqnarray}
F_{+1} &=& \pi K \log\left(\frac{R}{\xi}\right) + \frac{\pi K}{2}
\log\left(1-\frac{r_d^2}{R^2}\right) -4 \mu R
E\left(\frac{r_d}{R}\right) \nonumber \\
&&+ 2\pi\gamma R +\epsilon_{\rm core},
\label{F_interior}
\end{eqnarray}
where we have introduced a term for the core energy, $\epsilon_{\rm
core}$.  $E(x)$ is the complete elliptic integral of the second kind,
\begin{equation}
E(x) = \int_{0}^{\frac{\pi}{2}} \sqrt{1-x^2 \sin\phi} d\phi .
\end{equation}
We will see in section~\ref{sn:shapechanges}, when we investigate
changes in shape and texture, that when $r_d=0$ and $A_b<1$ this
texture is a local minimum with respect to all variations in the
texture (not just variations within this ansatz class).  However, we
find that $r_d=0$ and $A_b<1$ does not always yield a global minimum
even within this class (\ref{interior_texture}).
\par
\begin{figure}
\epsfxsize=3.375in
\centerline{\epsfbox{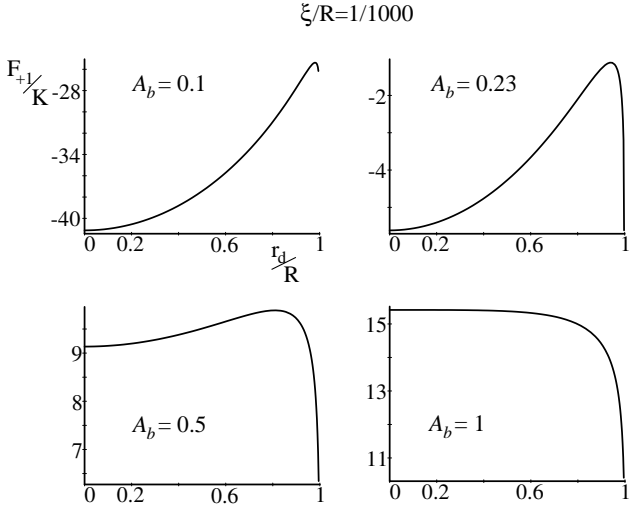}}
\caption{Plots of $F_{+1}/K$ vs $r_d / R$ for a disk of radius $R$ for several
values of $A_b$.  We have taken $\xi / R = 1/1000$ and have restricted
$r_d/R$ to be less than $1-1/1000$, so that the entire defect core is
within the disk.  We have also not included the core energy
$\epsilon_{\text core}$ or the isotropic line tension $2\pi\gamma\mu
R$ in $F_{+1}$, which here are merely uninteresting constants.  Note
that $r_d=0$ is a local minimum whenever $A_b<1$ and $r_d \approx R$
appears to be a local minimum for all of the plots.  When $A_b<0.23$
then $r_d=0$ appears to be the true minimum.  Notice that the barrier
separating the two minima is approximately $4 K$ at $A_b=0.23$. }
\label{fig_int_defect_stability}
\end{figure}

The free energy (\ref{F_interior}) has an interesting behavior as we
vary $r_d$.  As we move the defect off center, there is an energy
increase from the boundary term but a decrease in energy from the bulk
term.  The two contributions have quite different functional forms,
and we find that, depending upon the value of $A_b= K/ \mu R$, it may
be preferable either to keep the defect at the origin or to allow it
to migrate towards the boundary.  Maintaining $0<r_d<R-\xi$ we find
the following behavior of (\ref{F_interior}) within this class of
textures,
\begin{eqnarray}
A_b < 0.23 &\rightarrow \quad r_d=0 \text{ is a global minimum} \nonumber\\
A_b < 1 &\rightarrow \quad r_d=0 \text{ is a local minimum} \nonumber\\
A_b > 1 &\rightarrow \quad r_d=0 \text{ is a local maximum}. \label{AbTable}
\end{eqnarray}
This behavior is summarized in Fig.\ \ref{fig_int_defect_stability}.  We
see that within this class of textures there is a first-order
transition in the position of the defect as we change $A_b$.  When
$r_d=0$ becomes the global minimum, it may in fact already be the case
that the energy barrier that must be overcome in order for the defect
to migrate from the boundary to the origin may already be too high for
this to be a physically probable event.  Note that we have limited our
attention only to a particular subset of potential textures and this
barrier may not remain if we were to consider all possible textures.
However, we are aware of no experimental observations corresponding to
a single $+1$ defect near the boundary of the domain.  Thus, overall
we conclude that when a defect is present within the domain it is very
likely to be at the origin in a stable local minimum.

\begin{figure}
\epsfxsize=3.375in
\centerline{\epsfbox{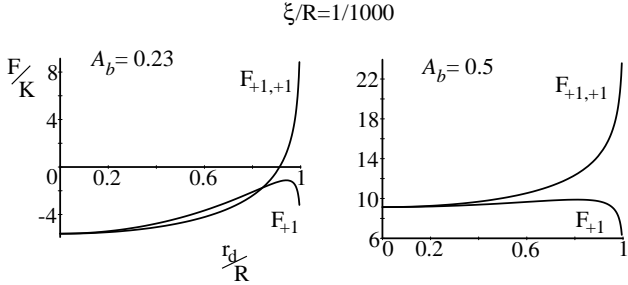}}
\caption{A comparison of the energy of an isolated $+1$ defect in the disk,
$F_{+1}$ with the energy of a $+1$ defect inside along with an image
defect outside enforcing strong pinning boundary conditions (${\bf \hat{m}}
\cdot {\bf \hat{t}} =1$), $F_{+1,+1}$.  Note that the energies are equal at
$R_d=0$.  At the boundary of the disk ($r_d=R-\xi$) $F_{+1}$ is always
lower.  Even though for some intermediate values of $r_d$ $F_{+1,+1}$
is smaller than $F_{+1}$ it is never substantially smaller. }
\label{fig_int_im_vs_noim}
\end{figure}

\par
We note that one can also consider the texture,
\begin{equation}
\Phi = \phi_{{\bf r}_d}({\bf r}) + \phi_{{\bf r}_d R^2/r_d^2}({\bf r}) - \frac{\pi}{2},
\end{equation}
which corresponds to a defect inside the domain along with an image
defect outside the domain which enforces strong pinning boundary
conditions, ${\bf \hat{m}} \cdot {\bf \hat{t}}=1$ (where again we take $\phi_d=0$).
Within this class textures the defect always prefers to be at the
origin, and when $A_b>0.23$ this texture provides at best a modest
improvement over (\ref{interior_texture}) and never offers a lower
global minimum (Fig.\ \ref{fig_int_im_vs_noim}).  Of course, when
$r_d=0$ this texture is the same as (\ref{interior_texture}) with
$r_d=0$.

\subsection{Which texture wins?} \label{sn:comparetextures}
\par
We have presented two textures which are local minima of the free
energy (\ref{Fmu}).  For the $+2$ exterior defect texture we found a
local minimum for all values of $A_b$ ( (\ref{exterior_texture}) with
$\phi_d=0$ and $r_d = r_d^0$ (\ref{ext_position}) ).  For the $+1$
interior defect texture we found a local minimum only for $A_b<1$
((\ref{interior_texture}) with ${\bf r}_d=0$).  Furthermore, we found that
this local minimum was not the global minimum even within the class of
ansatz textures (\ref{interior_texture}).  Here we will compare the
energies of the two locally minimal textures to see which is favored.
\par
For small $R$ the bulk energy dominates the boundary energy and we
expect the uniform texture to win.  Since the $+2$ defect texture
produces the uniform texture in the limit $A_b\rightarrow\infty$
($R\rightarrow 0$) we anticipate that for small $R$ this texture will
be favored.  For large $R$ ($A_b\rightarrow 0$) both textures satisfy
strong pinning boundary conditions and thus have equal contributions
from the surface energy.  However, the bulk energy cost of the
exterior defect texture is greater that that of the interior defect
texture.  This can be understood qualitatively by recalling that the
bulk energy cost of a defect of charge $q$ centered in a disk of
radius $R$ is $\pi K q^2 \log(R/\xi)$ \cite{Lubensky}.  In the limit
of large $R$ the $+1$ defect energy will be $\pi K \log(R/\xi)$
whereas the $+2$ defect energy will be $\approx (1/2)2^2 \pi K
\log(R/\xi)$.  The factor of $(1/2)$ arises because the $+2$ defect is
at the boundary of the disk, not the center.
\par
\begin{figure}
\epsfxsize=3.375in
\centerline{\epsfbox{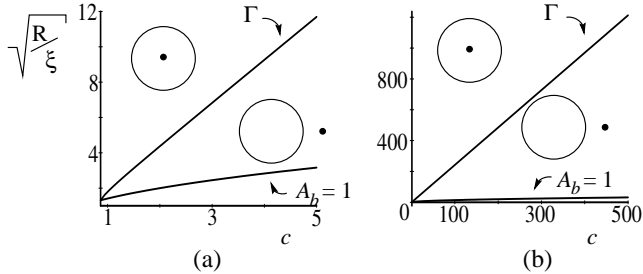}}
\caption{The two figures display at different scales the phase diagram
for the $+1$ defect at the center of the disk versus the exterior $+2$
defect texture.  The curve $\Gamma$ is $R=R_{\text{crit}}(c)$ where
$R_{\text{crit}}(c)$ is the critical radius for a given value of $c$
as discussed section~\ref{sn:comparetextures}.  The lower curve is
$A_b=1$.  To the left and above $\Gamma$ the interior defect texture
is favored, below the exterior defect is favored.  However, the
interior defect texture is a local minimum in the region above the
curve $A_b=1$ and the exterior region is a local minimum throughout
the region. }
\label{fig_phasediagram}
\end{figure}

Thus, we will find a first-order transition between the two textures.
In discussing this transition we find it more convenient to refer to
the dimensionless quantities
\begin{equation}
c = \frac{K}{2 \mu \xi} \quad \text{and} \quad x = \frac{R}{\xi}
\end{equation}
rather than $A_b$ and $R/\xi$.  Assuming that $R>\xi$ we find that
when $c<0.85$ the interior defect solution is favored.  For larger
values of $c$, there is a critical radius, $R_{\text{crit}}(c)$, above
which the interior defect texture is favored and below which the
exterior texture is favored.  Thus $R_{\text{crit}}(c)$ defines the
first-order phase boundary $\Gamma$ in Fig.\ \ref{fig_phasediagram}.
\par
Note in Fig.\ \ref{fig_phasediagram} that we have also included the
curve $A_b=1$ which is the boundary at which the interior defect
solution becomes unstable.  There is a large separation between this
boundary and the transition line $\Gamma$.  In the region above the
$A_b=1$ curve both textures are local minima and hence a physical
system may well find itself trapped in a meta-stable local minimum
when $A_b<1$.  Furthermore, in section~\ref{sn:defecttexture} we found
some circumstantial evidence suggesting that the barrier preventing
the expulsion of the interior defect may be quite large for physically
realisable values of $A_b$ (Fig.\ \ref{fig_int_defect_stability}).
Indeed in an experiment \cite{Knobler} that apparently has the
exterior defect texture, taking $\xi \approx 100$nm yields $c\approx
2$.  However, the radii of the domains appears to be as large as
$10\mu$m, which is much larger than the critical radius for $c=2$
($R_{\text{crit}}(2) \approx 2\mu$m), suggesting that it may be the
case that this texture is a meta-stable equilibrium state.

\par
To identify the transition we simply compare $F_{+1}$ and $F_{+2}$.
Taking $\epsilon_{\text{core}}=\pi K$ and noting that $A_b=2c/x$ we
can write (\ref{F_interior}) as
\begin{equation}
F_{+1} = \pi \mu R \left( \frac{2c}{x} \log x -2 + \frac{2c}{x} \right) + 2 \pi \gamma R.
\end{equation}
With $\phi_d=0$ and $r_d=r_d^0$ we can also write (\ref{F_exterior})
as
\begin{eqnarray}
F_{+2}&=& \pi \mu R \left( -\frac{4c}{x} \log 
         \left( 1- h^{2}\left(\frac{2c}{x}\right) \right)
         - 2h\left(\frac{2c}{x}\right) \right) \nonumber \\
         &&+ 2 \pi \gamma R.
\end{eqnarray}
where
\begin{equation}
h(y) = \frac{1}{y + \sqrt{1+y^2}}. 
\end{equation}
\par
Examining $F_{+1} - F_{+2}$ we find that when $c<c^* \approx 0.85$ the
interior defect solution is preferred for all values of $R$ (or
equivalently all values of $x$ since we are assuming that $\xi$ is
constant).  For larger values of $c$ we find that there is a critical
value of $R$, $R_{\text{crit}}(c)$, below which the exterior defect
solution is favored (Fig.\ \ref{fig_phasediagram}).

\section{Shape Changes} \label{sn:shapechanges}
\par 
For circular domains we have found two distinct textures which are
local minima of the free energy (this will be demonstrated soon).  Now
for each of these textures we would like to know if the configuration
($D, \Phi_0$), where $D$ is a disk and $\Phi_0$ is either of the two
textures, is a minimum with respect to changes in shape and texture.
We will find in each case that although the texture is stable when the
shape is fixed, the configuration (shape and texture) is unstable for
certain values of $A_b$ and $A_s$.  We will see that when $\mu>\gamma$
($A_s>0$) instabilities will begin to arise.
\par
To examine the stability of the configurations we need to introduce
perturbations in the shape and texture.  To this end we take $D$ to be
the region bounded by the curve, given in polar coordinates by
\begin{eqnarray}
r(\phi) &=& R_0 + \sum_{n \epsilon Z-\{0\}} \rho_n e^{in\phi} \nonumber \\
&=& R_0 + \sum_{n \epsilon Z^{+}} (a_n
\cos n\phi + b_n \sin n\phi),
\label{shape}
\end{eqnarray}
where
\begin{equation}
\rho_n = \rho_{-n}^* = \frac{1}{2}(a_n - ib_n) 
\end{equation}
and $R$ is defined by
\begin{eqnarray}
R_0 &=& R \left(1 - \sum_{n \epsilon Z-\{0\}} \frac{\rho_n
              \rho_{-n}}{R^2}\right)^{\frac{1}{2}} \\ 
      &=& R \left(1 - \frac{1}{2} \sum_{n \epsilon Z^{+}} \frac{a_n^2 +
      b_n^2}{R^2}\right)^{\frac{1}{2}}. \nonumber
\end{eqnarray}  
With this parameterization the area of the domain is simply
\begin{equation}
A = \pi R^2.
\end{equation}
Thus keeping $R$ fixed keeps the area constant.  Note that we have
included the $n=1$ terms in (\ref{shape}).  Although these modes can
usually be ignored as being just translations of $D$ (to lowest order
in the Fourier coefficients), here the texture distinguishes a
particular origin for the coordinate system and as such translations
of $D$ are associated with nontrivial changes in the configuration
(see App. \ref{shapeapp}).
\par
To allow for perturbations in the texture we will take
\begin{equation}
\Phi({\bf r}) = \Phi_0({\bf r}) + \sum_{n \epsilon Z-\{0\}} r^{|n|}P_ne^{in\phi} \quad
\text{with} \quad P_n=P_{-n}^{*},
\label{texture} 
\end{equation}
where $\Phi_0({\bf r})$ is one of the two textures found in
section~\ref{sn:texture} ( (\ref{exterior_texture}) and
(\ref{interior_texture})) and the perturbation is the most general
nonsingular addition allowed by the Euler-Lagrange equation $\nabla^2
\Phi =0$.  Taking $O(P_n) = O(\rho_n) = O(\epsilon)$ we will proceed
to calculate $F$ to order $\epsilon^2$.  This will provide us with an
approximation to
\begin{equation}
F(\bbox{\rho}, {\bf P}, K, \gamma, \mu, R),
\end{equation}
where
\begin{eqnarray}
&&\bbox{\rho} = (\rho_1,\rho_{-1},\rho_2, \rho_{-2}, \ldots) \quad \text{and}
\nonumber \\
&&{\bf P} = (P_1,P_{-1},P_2,P_{-2}, \ldots).
\end{eqnarray}
Minimizing this over ${\bf P}$ we will obtain the effective free energy,
\begin{equation}
F_{\text{eff}}(\bbox{\rho}, K, \gamma, \mu, R),
\end{equation}
as a function of the elastic constants, the radius, and the shape
perturbation parameters correct to $O(\epsilon^2)$.
\par
In both cases we find that $F_{\text{eff}}$ contains no terms of
$O(\epsilon)$, that is no terms linear in ${\rho_n}$, demonstrating
that the configurations are extremal.  We also find that when
$\mu>\gamma$ shape instabilities will be present for certain values of
$A_b$ and $A_s$.  In fact, for any given $A_s>0$ we find that
instabilities arise for sufficiently small $A_b$'s.
\par
Ideally we would like to calculate $F_{\text{eff}}$ correct to all
orders in $\{\rho_n\}$.  Then minimization of $F_{\text{eff}}$ over
the $\rho_n$'s would yield distorted equilibrium shapes.
Unfortunately this is a prohibitively difficult task.  Alternatively,
we would like to calculate $F_{\text{eff}}$ to high enough order in
$\epsilon$ such that the effective energy is bounded below.  This
would provide approximations to the equilibrium values of the
$\rho_n$'s, and thus approximations to distorted equilibrium domain
shapes.  Calculating $F_{\text{eff}}$ to $O(\epsilon^4)$, while
difficult but not impossible, is not sufficient.  This appears to be
due to the form of $\gamma
\oint_{\partial D} ds$ as a function of the $\rho_n$'s.  Although this
isotropic line tension term should be stabilizing the boundary against
deformations, the fourth order coefficients are negative and as such
they lead to $F_{\text{eff}}$ being unbounded below at $O(\epsilon^4)$.
Because of extensive mode coupling and the exponential increase in the
number of terms in $F_{\text{eff}}$ we have not proceeded to
$O(\epsilon^6)$.

\subsection{The exterior defect} \label{sn:extshape}
\par
We will now apply the outlined procedure to the case of the exterior
defect.  We take $D$ as in (\ref{shape}) and we take
\begin{equation}
\Phi_0 = 2 \phi_{{\bf r}_d^0}({\bf r}) - \frac{\pi}{2},
\end{equation}
the preferred texture for a disk, as found in
section~\ref{sn:boojumtexture}. We will not allow ${\bf r}_d^0$ to vary,
nor will we allow for variations in the constant term ($C=-\pi / 2$).
Allowing ${\bf r}_d^0$ to vary would present us with a redundancy in the
definition of $\Phi$ (\ref{texture}).  Namely, changes in ${\bf r}_d^0$
are already accounted for by the $P_n$'s in (\ref{texture}).  If we
wished to have
\begin{equation}
\Phi = 2\phi_{{\bf r}_d} - \frac{\pi}{2}
\end{equation}
where ${\bf r}_d \neq {\bf r}_d^0$ this could be achieved by an appropriate
choice of the $P_n$'s in (\ref{texture}) since the perturbative terms
form a complete set of solutions to Laplace's equation in the disk.
\par
Noting the $\sin(C+\phi_d)$ term in (\ref{F_exterior_phid}), we see
that rotating the defect about the center of the disk and
simultaneously adjusting $C$ to maintain $C+\phi_d= -\pi / 2$ costs no
energy.  As we have just noted, an appropriate choice of the $P_n$'s
will produce this rotation, and thus if $C$ were allowed to vary, we
would discover this Goldstone mode eventually.  By fixing $C$ to be
$-\pi / 2$ we freeze out this mode.  Experimental evidence suggests
that this is the appropriate course to take.  The texture associated
with a $+2$ virtual defect has been observed \cite{Riviere95,Knobler},
demonstrating that on the time scale of the observations the virtual
defect does not freely rotate about the domain.  If the defect did
execute such a motion, then the time averaged texture would appear as
the trivial texture.  Though a careful consideration of the dynamics
might be in order, in this paper we will not concern ourselves further
with the origin of the observed angular stability of the virtual
defect.
\par
We find it more convenient now to use the real rather than the complex
fourier coefficients in calculating $F$:
\begin{eqnarray}
A_n = 2{\rm Re}[P_n] \quad B_n=2{\rm Im}[P_n] \quad &&P_n =\frac{1}{2}(A_n+iB_n)
\nonumber\\
a_n = 2{\rm Re}[\rho_n] \quad b_n =2{\rm Im}[\rho_n]  \quad
&&\rho_n =\frac{1}{2}(a_n+ib_n). 
\end{eqnarray}
Thus, now ${\bf A}=(A_1,A_2,\ldots)$ and ${\bf B}=(B_1,B_2,\ldots)$ describe
texture changes whereas ${\bf a}=(a_1,a_2,\ldots)$ and
${\bf b}=(b_1,b_2,\ldots)$ describe shape changes.
Taking $K=1$ and $R=1$ to set our energy and length scales, and
performing a plethora of contour integrals we eventually find
\begin{eqnarray}
F({\bf A},{\bf B},{\bf a},{\bf b},A_b,A_s) &=& \langle{\bf A} |H_{AA}| {\bf A} \rangle
                            + \langle{\bf B} |H_{BB}| {\bf B} \rangle \nonumber\\
                           &+& \langle{\bf a} |H_{aa}| {\bf a} \rangle
                            + \langle{\bf b} |H_{bb}| {\bf b} \rangle \nonumber \\
                           &+& \langle{\bf A} |U_{Ab}| {\bf b} \rangle
                            + \langle{\bf B} |U_{Ba}| {\bf a} \rangle  \label{FABab}\\
                &+& F_{+2}, \nonumber
\end{eqnarray}
where
\begin{eqnarray}
&&(H_{AA})_{n,m}=(H_{BB})_{n,m} = \\
&& \frac{-\mu \pi}{4}
 \left( \frac{1}{(r_d^0)^{m+n-1}} - \frac{1}{(r_d^0)^{m+n+1}}\right)
\nonumber \\
&& -\frac{\mu \pi}{4}
 \left(\frac{1- \delta_{m,n}}{(r_d^0)^{|m-n|-1}} 
       - \frac{1- \delta_{m,n}}{(r_d^0)^{|m-n|+1}} \right) \nonumber \\
 && + \frac{\mu \pi}{2 r_d^0}\delta_{m,n}
 + 2\pi m\delta_{m,n}, \nonumber
\end{eqnarray}

\begin{eqnarray}
&&(U_{Ab})_{n,m} = \\
&&\frac{\mu \pi}{r_d^0} 
 \left( \frac{1}{(r_d^0)^{m+n}} + 
   (-1)^{n-m+1}2\frac{1-\delta_{m,n}}{(r_d^0)^{|n-m|}}
  -2 m \delta_{m,n} \right), \nonumber
\end{eqnarray}

\begin{eqnarray}
&&(U_{Ba})_{n,m} = \\
&&\frac{\mu \pi}{r_d^0} 
 \left( \frac{1}{(r_d^0)^{m+n}} + 
   (-1)^{n-m}2\frac{1-\delta_{m,n}}{(r_d^0)^{|n-m|}}
  + 2 m \delta_{m,n} \right), \nonumber
\end{eqnarray} 

and
\begin{equation}
(H_{aa})_{n,m} = \frac{1}{2} (h(n,m) + h(n,-m)) 
            + \frac{1}{2}\pi\gamma (n^2-1)\delta_{m,n},
\end{equation}

\begin{equation}
(H_{bb})_{n,m} = \frac{1}{2} (-h(n,m) + h(n,-m)) 
            + \frac{1}{2}\pi\gamma (n^2-1)\delta_{m,n},
\end{equation}
where
\begin{eqnarray}
&&h(n,m) = \pi \mu (1-\delta_{m,-n}) \\
&&\times \Bigg[ \frac{|m+n|-2}{2}
 \left(\frac{|m+n|+1}{(r_d^0)^{|m+n|+1}}
      - \frac{|m+n|-1}{(r_d^0)^{|m+n|-1}}\right) \nonumber \\
&&+ \frac{(|m+n|-2)(|m+n|-1)}{2(r_d^0)^{|m+n|-1}} - 
        \frac{|m+n|(|m+n|+1)}{2(r_d^0)^{|m+n|+1}}
      \Bigg] \nonumber \\
&&- \frac{2\pi}{(r_d^0)^2-1}\delta_{m,-n} \nonumber \\
&&+ \frac{2\pi}{(r_d^0)^{|m+n|}}\left(\frac{|m+n|+1}{(r_d^0)^2-1}
     +\frac{2}{((r_d^0)^2-1)^2} \right). \nonumber
\end{eqnarray}

For convenience we have chosen to leave the above in terms of $\mu$,
$\gamma$ and $r_d^0$.  With $K=1$ and $R=1$ we have, 
\begin{eqnarray}
\mu &=& \frac{1}{A_b} \nonumber \\
r_d^0 &=& A_b + \sqrt{A_b^2+1} \\
\gamma &=& \frac{\mu}{A_s+1} = \frac{A_b}{A_s+1}. \nonumber
\end{eqnarray}
\par
Note the absence of any terms linear in $\{{\bf A},{\bf B},{\bf a},{\bf b}\}$ in
(\ref{FABab}).  This confirms that the unperturbed configuration is an
extremum.  Furthermore, $H_{AA}, H_{BB}, H_{aa}$ and $H_{bb}$ are all
positive definite.  Thus the unperturbed configuration is actually a
minimum with respect to changes in either texture or shape alone.  Any
instabilities must therefore arise as a result of the couplings,
$U_{Ab}$ and $U_{Ba}$, between shape and texture.
\par
Before minimizing over ${\bf A}$ and ${\bf B}$ to find $F_{\text{eff}}$ we note that
two Goldstone modes are still present.  A simultaneous translation of
the domain and the defect will not change the energy, $F$.  This
implies the existence of 2 zero eigenvalues in $F$, one for horizontal
and one for vertical translations.  These translations are achieved
through appropriate choices for ${\bf A},{\bf B},{\bf a}$ and ${\bf b}$.  For example,
the infinitesimal translation $\epsilon {\bf \hat{e}}_x$ is achieved by taking
\begin{eqnarray}
a_1 &=& \epsilon \nonumber \\
B_n &=& -\frac{2\epsilon}{(r_d^0)^{n+1}}.
\end{eqnarray}
The two Goldstone modes should be present in $F_{\text{eff}}$ and we will use
their existence as a check on our calculations.
\par
Symbolically, calculating $F_{\text{eff}}$ is straightforward,
\begin{equation}
F_{\text{eff}}({\bf a},{\bf b}) = \langle{\bf a}|H_{aa}+Q_{aa}|{\bf a}\rangle
                 + \langle{\bf b}|H_{bb}+Q_{bb}|{\bf b}\rangle,
\end{equation}
where
\begin{eqnarray}
Q_{aa} &=& -\frac{1}{4}U_{Ba}^{T}H_{BB}^{-1}U_{Ba} \nonumber\\
Q_{bb} &=& -\frac{1}{4}U_{Ab}^{T}H_{AA}^{-1}U_{Ab}. 
\end{eqnarray}
However, the actual calculation of $F_{\text{eff}}$ is not so easy.  Neither
$H_{AA}$ nor $H_{BB}$ is diagonal, the off-diagonal terms are not in
general small, and the higher order harmonics (in ${\bf A}$ and ${\bf B}$) are
not necessarily unimportant.  We will resort to numerical procedures
to aid us in finding the eigenvalues and eigenvectors of $F_{\text{eff}}$.
\par
If $r_d^0$ is large then the higher harmonics in ${\bf A}$ and ${\bf B}$ do
become less important, and we anticipate that working with finite
dimensional $H_{AA}$'s and $H_{BB}$'s will provide a reasonable
approximation to $F_{\text{eff}}$.  Furthermore, if $\gamma \neq 0$ then the
higher harmonics in ${\bf a}$ and ${\bf b}$ are also rapidly suppressed in
$H_{aa}$ and $H_{bb}$, allowing us to work with finite dimensional
$H_{aa}$'s, $H_{bb}$'s and $U$'s as well.
\par
\begin{figure}
\epsfxsize=3.375in
\centerline{\epsfbox{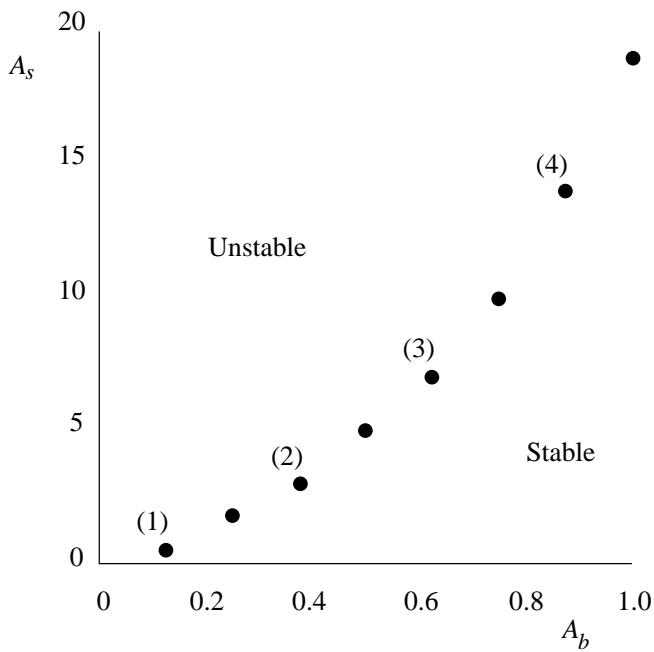}}
\caption{The stability boundary for ($H_{aa}+Q_{aa}$) and ($H_{bb}+Q_{bb}$) as
found by using 12 texture and 6 shape modes.  The two boundaries are
apparently the same within the accuracy of the calculation ($\approx
10\%$).  The unstable eigenmodes associated with the noted points are
given in the text, one mode each for ($H_{aa}+Q_{aa}$) and
($H_{bb}+Q_{bb}$).  Above these points on the boundary the unstable
modes appear to be pure $n=2$ modes.  As noted in the text, it seems
likely that the true unstable modes are always the $n=2$ modes.  }
\label{fig_ext_shape_stability}
\end{figure}

Noting that ${\bf a}=(1,{\bf 0})$ and ${\bf b} = (1,{\bf 0})$ are the
Goldstone modes for $F_{\text{eff}}$ provides us with one gauge of how
well we have approximated $F_{\text{eff}}$ with finite dimensional
matrices.  Even if we do include too few modes to correctly identify
these Goldstone modes, we will still find bounds on the stability.
Including more modes will only make the system more unstable.
Fig.\ \ref{fig_ext_shape_stability} shows the stability boundary for
$H_{aa}+Q_{aa}$ and $H_{bb}+Q_{bb}$ in the $(A_b,A_s)$ plane.  We have
used 12 texture modes and 6 shape modes in the numerical calculations.
Below are the unstable eigenmodes at the stability boundary for the
points indicated in Fig.\ \ref{fig_ext_shape_stability},
\begin{eqnarray}
(1) \quad &{\bf a}& \approx (0.70, -0.71, 0.01, {\bf 0}) \nonumber\\
&{\bf b}& \approx (0.70, -0.71, 0.01, {\bf 0}) \nonumber\\
(2) \quad &{\bf a}& \approx (0.04, -0.99, 0.01, {\bf 0}) \nonumber\\
&{\bf b}& \approx (0.04, -0.99, 0.01, {\bf 0}) \\
(3) \quad &{\bf a}& \approx (0, 1.0, {\bf 0}) \nonumber\\
&{\bf b}& \approx (0, 1.0, {\bf 0}) \nonumber\\
(4) \quad &{\bf a}& \approx (0, 1.0, {\bf 0}) \nonumber\\
&{\bf b}& \approx (0, 1.0, {\bf 0}) \nonumber
\end{eqnarray}
\par
We see that when $A_s$ is small the first unstable mode contains
significant contributions from the $n=1$ mode indicating that the
Goldstone modes ${\bf a}=(1,{\bf 0})$ and ${\bf b}=(1,{\bf 0})$ have not been
properly identified within the numerical approximation.  For larger
values of $A_s$ the first unstable modes are pure $n=2$ modes and the
numerical calculation also correctly identifies the $n=1$ modes as
zero-eigenvalue mode.  Now looking back at the eigenmodes for small
$A_s$ we note that only the $n=1$ and $n=2$ modes contribute
significantly, suggesting that the pure $n=2$ mode is always the first
mode to go unstable.

\par
Fig.\ \ref{fig_ext_shape_stability} shows us that for any $A_s>0$ there is
a critical value of $A_b$ below which the system becomes unstable.
Though not shown in the figure, we remark that as $A_b$ becomes even
smaller, more and more modes become unstable.  This should not be
surprising since, if $A_b=0$ and $A_s>0$ then all modes are unstable.
\par
Thus we find that for the free energy (\ref{Fmu}) the configuration
consisting of the disk with the preferred exterior defect texture [
(\ref{exterior_texture}) with $\phi_d=0$ and $r_d=r_d^0$
(\ref{ext_position})] is unstable with respect to correlated changes
of the shape and texture when $A_s>0$.  Furthermore, the absence of
linear terms in ${\bf a}$ and ${\bf b}$ in $F$ arises from a
cancellation between the bulk and boundary portions of the free
energy.  Hence, changing the structure of either will generically
generate such linear terms leading to linear shape instabilities.
This helps us understand why either altering the form of the boundary
energy or accounting for deviations from the one-coupling-constant
approximation also leads to shape instabilities.

\subsection{The interior defect} \label{sn:intshape}
\par
Now we will examine the situation with the interior defect.  Again, we
take $D$ as in (\ref{shape}) and now we take
\begin{equation}
\Phi_0 = \phi + \frac{\pi}{2}, 
\end{equation}
the preferred texture for a disk as found in
section~\ref{sn:defecttexture}.  This time, since the defect is inside
the disk, it is not possible to make an appropriate choice of the
$P_n$'s in (\ref{texture}) to change the position of the defect.
Nevertheless, we will not allow the defect location to vary.  This
merely has the effect of removing the Goldstone modes associated with
simultaneous translations of the texture and the shape.
\par
We will again work only to quadratic order in $\bbox{\rho}$ and ${\bf P}$ and
extract an effective free energy $F_{\text{eff}}(\bbox{\rho},K,\gamma,\mu,R)$.
As with the exterior defect calculation this $F_{\text{eff}}$ will
contain no linear terms in $\bbox{\rho}$ and the quadratic terms will not
always be positive.  We will be able to investigate analytically the
stability criteria in this case, and we will find again that for any
$A_s>0$ and for small enough $A_b$ shape instabilities arise.  The
answer to the question of which mode becomes unstable first will be
much more interesting here.  We will find that for the $k$th harmonic
mode there is always a value of $A_s$ such that this mode is the first
to go unstable as $A_b$ is lowered from $1$.  It will always be
assumed that $A_b<1$ since for larger values we already know from
section~\ref{sn:defecttexture} that the defect would prefer to move away
from the origin.  This will appear in our current analysis as an
instability in the $n=1$ mode.
\par
As with the exterior defect case, here again it would be necessary to
proceed to sixth order in $\bbox{\rho}$ in the calculation of $F_{\text{eff}}$ in
order to calculate equilibrium shapes.  Again this task appears
formidable and we will only be examining the stability of the
configurations. 
\par
Proceeding, we have
\begin{eqnarray}
F &=&\pi K \log\left(\frac{R}{\xi}\right) - \pi K \sum_{n \epsilon Z-\{0\}}
\frac{\rho_n \rho_{-n}}{R^2} \nonumber \\
&&+ 2\pi K \sum_{n \epsilon Z-\{0\}} i n P_n R^{|n|} \frac{\rho_{-n}}{R} + 2 \pi K
\sum_{n=1} n P_n P_{-n} R^{2n} \nonumber \\
&&- 2\pi \mu R + \pi \mu R \sum_{n \epsilon Z-\{0\}} \frac{\rho_n  \rho_{-n}}{R^2}
\nonumber \\
&&+ \pi\mu R \sum_{n \epsilon Z-\{0\}}R^{2|n|} P_n P_{-n} \nonumber \\
&& - 2\pi i \mu R \sum_{n \epsilon Z-\{0\}} n R^{|n|} P_n
\frac{\rho_{-n}}{R} \nonumber \\ 
&&+ 2 \pi \gamma R + \pi \gamma R \sum_{n \epsilon Z-\{0\}} (n^2-1) \frac{\rho_n
\rho_{-n}}{R^2}. 
\label{Finterior}
\end{eqnarray}
The absence of any terms linear in ${\bf P}$ with no $\bbox{\rho}$'s confirms the
extremal nature of the texture.  The positive definiteness of the
quadratic form associated with the ${\bf P}$'s confirms that the texture
is in fact a minimum with respect to variations that do not change the
position of the defect.

\begin{figure}[t]
\twocolumn[\hsize\textwidth\columnwidth\hsize\csname@twocolumnfalse\endcsname
\epsfxsize=7in
\centerline{\epsfbox{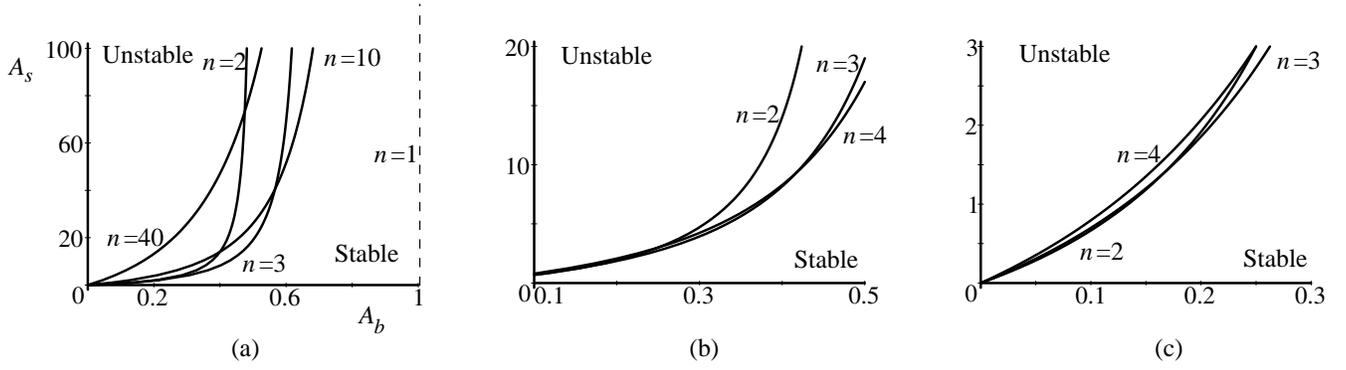}}
\caption{Stability boundaries in the $(A_b,A_s)$-plane for a few modes.  Note
the $n=1$ boundary in (a), this mode is the first to go unstable as
$A_b$ increases beyond $1$, instabilities in the other modes for
$A_b>1$ are not shown.  This instability corresponds to the defect
wishing to move away from the origin.  In all three pictures note the
crossing of the stability boundaries.  Starting in the stable region,
as $A_b$ is lowered which mode goes unstable first clearly depends
upon the value of $A_s$.  For completeness we note that the $n$th
stability curve has an asymptote at $A_b=1-1/n$ as one can see in (a)
for the $n=2,3$ curves. }
\label{fig_int_shape_stability}
\vskip 2pc]
\end{figure}

\par
To calculate $F_{\text{eff}}$ we simply demand that $\partial F / \partial
P_j =0$ and solve for  $P_j({\rho_n})$:
\begin{eqnarray}
\frac{\partial F}{\partial P_j} &=& 2 \pi\left( K i j R^|j|
\frac{\rho_{-j}}{R} + K |j| P_{-j}R^{2|j|} \right) \nonumber \\
&+& 2\pi \left( \mu R R^{2|j|} P_{-j} -
\mu i j R R^{|j|} \frac{\rho_{-j}}{R} \right)
\end{eqnarray}
\begin{equation}
\text{yields} \qquad P_j = i \frac{j}{|j|} \delta_j \frac{1}{R^{|j|}} \frac{\rho_j}{R}
\end{equation}
\begin{equation}
\text{where} \qquad \delta_j = \frac{K - \mu R}{K + \frac{1}{|j|}
\mu R}. 
\end{equation}
Substituting into (\ref{Finterior}) we have
\begin{eqnarray}
F_{\text{eff}} &=& \pi K \log \left(\frac{R}{\xi}\right) -2 \pi \mu R +
2 \pi \gamma R \nonumber \\
&&+ 2 \pi \mu R \sum_{n=1} \Omega_n \frac{\rho_n \rho_{-n}}{R^2} + O(\bbox{\rho}^3)
\end{eqnarray}
where
\begin{equation}
\Omega_n =  - \left( A_b-1+n\frac{(A_b-1)^2}{A_b+\frac{1}{n}} -
\frac{1}{1+A_s} (n^2-1) \right).
\end{equation}
Note that to $O(\bbox{\rho}^2)$ $F_{\text{eff}}$ is block diagonal in the
$\rho_n$'s and diagonal in the corresponding $a_n$'s and $b_n$'s.  We
can examine the stability of the circle simply by looking at the sign
of $\Omega_n$.  If $\Omega_n$ is negative then the $n$th harmonic is
unstable.
\par
The stability of the $n=1$ mode is independent of $A_s$ and we can
easily see that this mode is unstable when $A_b>1$.  This corresponds
precisely to our earlier finding (\ref{AbTable}) that the origin
becomes a local maximum for the position of the defect when $A_b>1$.
\par
For the other modes the boundary between the stable and unstable
regions in the $(A_b,A_s)$-plane is more complex.  For the $n$th mode
this boundary is given by the curve $\Omega_n=0$, or equivalently by,
\begin{equation}
A_s^0(n,A_b) = -\frac{A_b}{1-A_b}\frac{n^2+n-1-nA_b}{nA_b-n+1}.
\end{equation}
We are only interested in the region $(A_s>-1 , 0<A_b<1)$, since
physically we require $(A_b>0, A_s>-1)$ and we have already seen that
the $n=1$ mode becomes unstable when $A_b>1$.  In
Fig.\ \ref{fig_int_shape_stability} we have plotted $A_s^0(n,A_b)$ in
this region for several values of $n$.  Within this piece of the
$(A_b,A_s)$-plane the $n$th harmonic mode is stable in the region to
the right of $A_s^0(n,A_b)$ and unstable to the left of this curve.
\par
Fixing the physical elastic constants of the system and allowing the
size of the domain ($R$) to change is, in the $(A_b,A_s)$-plane,
equivalent to fixing $A_s$ and allowing $A_b$ to vary.  Thus, a
growing domain traces out a horizontal line in the $(A_b,A_s)$-plane.
Starting in the stable region (to the right of the stability curves),
as the domain grows which stability boundary the horizontal line
crosses first will clearly depend upon the value of $A_s$.  In fact
for 
\begin{equation}
\frac{m^4-4m^3+5m^2-4m+1}{2m -1} <A_s< \frac{m^4-m^2-2m-1}{2m+1}
\end{equation}
the stability boundary for the $m$th harmonic mode will be the first
such boundary crossed.  Thus the domain will become unstable with
respect to $m$-fold shape perturbations first.  Of course, as the
domain grows further it will become unstable with respect to more and
more modes.  Furthermore there is extensive mode coupling at higher
order in $\bbox{\rho}$, and as such the equilibrium shape may not exhibit
$m$-fold shape distortions.  Nevertheless, this behavior is markedly
different from what we found in the case of the exterior defect.
There it appeared that the $n=2$ mode always became unstable first.
Interestingly, in the extensively investigated dipole model, the $n=2$
mode also appears to be the first mode to become unstable.  While in
both of these cases higher harmonics also become unstable as the
domain grows larger, and mode coupling is present, it is still a
possibility that $m$-fold shape distortions may be more easily
accounted for by the present model with a captured defect.

\section{Conclusions} \label{sn:Conclusions}
\par
After a brief review of the relevant types of order present in domains
appearing in the coexistence regions of mono-layer films we focused on
the effects of the tilt order through the simple $XY$-model.  We have
shown the possible existence of a first-order transition in the
texture for a circular domain from a virtual defect texture to a
captured defect texture.  We then examined the shape instabilities for
both of these textures.  Through a slightly different parameterization
of the domain we were able to reproduce the result of Rudnick and
Bruinsma \cite{Bruinsma} that the circle is an extremum for the
virtual defect texture.  However, we find that it is not always a
minimum and shape instabilities can arise whenever there is the
possibility, locally, of a negative effective line tension $(A_s>0)$.
For the interior defect texture we have found a complex stability
landscape for the fourier modes associated with the domain shape,
again arising when $A_s>0$.  In both cases larger domains become
unstable with respect to the fourier modes.  With the exterior defect
texture, the $n=2$ mode apparently is always the first to become
unstable.  With the interior defect texture which mode becomes
unstable first depends upon the physical parameters of the system.
\par
While the dipole model can also have shape instabilities in its
fourier modes, it cannot produce chiral domain shapes.  However, for
unequal elastic constants the $XY$-model may yield chiral shapes.  Of
course, a general system may have important contributions from the
dipole model and from the $XY$-model.

\acknowledgements
We are grateful for helpful conversations with R. Bruinsma, R. Kamien,
and J. Rudnick.  This work was supported primarily by the Materials
Science and Engineering Center Program of NSF under award DMR96-32598.

\appendix
\section{Shape Parameterization} \label{shapeapp}
\par
Here we review some of the properties, including advantages and
shortcomings, of our choice of parameterization for the boundary of
the domain $D$.  Recall (\ref{shape})
\begin{equation}
r(\phi) = R_0 + \sum_{n \epsilon Z^{+}} (a_n\cos n\phi + b_n \sin n\phi).
\label{shape_app}
\end{equation}
Next recall that a domain, $D$, is said to be star-shaped if there
exists a point $P_0 \epsilon D$ such that, for all $P \epsilon D$ the
line segment connecting $P_0$ to $P$ is contained in $D$.  The
boundary of any star-shaped domain, where the origin can serve as the
point $P_0 \epsilon D$, can be described by (\ref{shape_app}).
Correspondingly, a curve given by (\ref{shape_app}), if it is also a
bounded simple closed curve, serves as the boundary of a star-shaped
domain where the origin can be chosen as $P_0$.
\par
\begin{figure}
\epsfxsize=3.375in
\centerline{\epsfbox{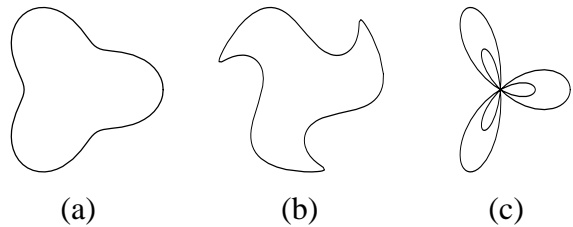}}
\caption{The simple closed curve in (a) is admitted by the parameterization
(\ref{shape_app}), it bounds a star-shaped domain.  In (b) this simple
closed curve does not bound a star-shaped domain and is not admitted
by the parameterization (\ref{shape_app}).  Finally in (c) we have a
closed curve with self-intersections which is admitted by
(\ref{shape_app}).     
 }
\label{fig_appendix}
\end{figure}

A priori we have no reason to restrict our attention to star-shaped
domains.  Unfortunately our parameterization does not allow the
description of domains that are not star-shaped and that have
boundaries that are simple closed curves (Fig.\ \ref{fig_appendix}).
Furthermore, the parameterization (\ref{shape_app}) admits curves that
are not simple closed curves and that we would consider to be
physically irrelevant (Fig.\ \ref{fig_appendix}).  Thus, if we were
to attempt to find highly distorted equilibrium domain shapes, we
would be well advised to use a different parameterization.

\par
In this paper, however, we are only interested in investigating the
stability of a circular domain.  This can be accomplished with the
parameterization (\ref{shape_app}) since the circular domain is
star-shaped and physically relevant infinitesimal perturbations will
also be star-shaped.
\par
Another potential difficulty with this parameterization is the special
role the origin assumes.  Namely, we must be allowed to choose the
origin as our special point $P_0$.  Thus, for example, a circle of
radius $1$ centered about $(2,0)$ cannot be described by the
parameterization (\ref{shape_app}).
\par
If we are only concerned with the {\em shape} of the domain, then, if
it is star-shaped, we can always choose the origin to be a point in the
domain satisfying the properties of $P_0$.  But now if we are not
concerned with the position of the domain with respect to the origin
of the coordinate system, then our parameterization contains
redundancies.  Namely, for each point in the domain that satisfies
the required properties of $P_0$, we can find a parameterization of the
form (\ref{shape_app}) which produces the correct boundary.  For
example, for a disk any interior point could be chosen as the origin.
Translating a domain around in general involves changing all of the
fourier coefficients and so removing this redundancy is not
necessarily trivial.  However, again, being interested only in small
perturbations of a circular domain allows us to easily deal with this
redundancy.  Starting with a circular boundary centered at the origin,
\begin{equation}
r(\phi) = R,
\end{equation}
one finds that for an infinitesimal translation, $(\epsilon\cos\psi,
\epsilon\sin\psi)$, the translated boundary is given by,
\begin{equation}
r(\phi) = R + (\epsilon\cos\psi)\cos\phi + (\epsilon\sin\psi)\sin\phi
+ O(\epsilon^2).
\end{equation}
That is, to $O(\epsilon)$ only the $n=1$ modes are altered.  It is for
this reason that one often ignores the $n=1$ modes when using this
parameterization.  To lowest order these modes are merely translations
of the domain, and as such merely produce annoying redundancies in the
parameterization if one is interested only in the {\em shape} of the
domain.  However, in our problems we also have a nontrivial texture
present which breaks the isotropy of space.  Thus the position of the
domain relative to a fixed origin (provided by the texture) is now
physically relevant.  The redundancies are now in the simultaneous
translations of the texture field and the domain.

\section{The choice of $\eta=0$} \label{nochiralapp}
\par
We would like to comment here on the invariance of our results with
respect to our choice of boundary energy, namely $\eta=0$ in
(\ref{Fmu}).
\par
Let's consider the two free energies,
\begin{eqnarray}
F_{\text{c}} &=& \frac{1}{2} K \int_D (\nabla \Phi)^2 d^2x
+\eta\oint_{\partial D} {\bf \hat{m}} \cdot {\bf \hat{n}} ds \nonumber \\ 
&&-\mu\oint_{\partial D} {\bf \hat{m}} \cdot {\bf \hat{t}} ds 
+ \gamma \oint_{\partial D} ds
\end{eqnarray}
and
\begin{equation}
F_{\text{a}} = \frac{1}{2} K \int_D (\nabla \Phi^{'})^2 d^2x
+\eta^{'} \oint_{\partial D} {\bf \hat{m}}^{'} \cdot {\bf \hat{n}} ds 
+ \gamma \oint_{\partial D} ds.
\end{equation}
Under the transformation
\begin{eqnarray}
&&\Phi' = \Phi' + \tan^{-1}\frac{\mu}{\eta} \nonumber \\
&&\eta' = \sqrt{\mu^{2} + \eta^{2}}
\end{eqnarray}
we see that $F_{\text{c}} = F_{\text{a}}$.  Thus, it is trivial to
extend our results for $\eta=0$ to arbitrary values of $\eta$ and
$\mu$.
\par
Interestingly the domain $D$ does not participate in the
transformation.  Thus, although $F_{\text{c}}$ contains a chiral term,
indicating that the equilibrium configuration $(D,{\bf \hat{m}})$ should be
chiral, we expect $D$ to be achiral.  This follows from the
expectation that the equilibrium configuration for $F_{\text{a}}$,
$(D,{\bf \hat{m}}')$, should be achiral since $F_{\text{a}}$ contains no
chiral terms.  Thus, the chirality only manifests itself in the
equilibrium configuration through the texture ${\bf \hat{m}}$ and not through
the shape $D$.

\end{document}